# Advances in quantum metrology with dielectrically structured single photon sources based on molecules

*Pietro Lombardi\*, Hristina Georgieva, Franziska Hirt, Juergen Mony, Rocco Duquennoy, Ramin Emadi, Maria Guadalupe Aparicio, Maja Colautti, Marco López, Stefan Kück and Costanza Toninelli\**

Pietro Lombardi

Istituto Nazionele di Ottica (CNR–INO), c/o European Laboratory for Non-Linear Spectroscopy (LENS), Via Nello Carrara 1, Sesto Fiorentino, 50019, Italy

European Laboratory for Non-Linear Spectroscopy (LENS), Via Nello Carrara 1, Sesto Fiorentino, 50019, Italy

E-mail: pietroernesto.lombardi@ino.cnr.it

Hristina Georgieva

Physikalisch-Technische Bundesanstalt (PTB), Bundesallee 100, 38116 Braunschweig, Germany

Franziska Hirt

Physikalisch-Technische Bundesanstalt (PTB), Bundesallee 100, 38116 Braunschweig, Germany

Laboratory for Emerging Nanometrology, Langer Kamp 6a/b, 38106 Braunschweig, Germany

Juergen Mony

Istituto Nazionele di Ottica (CNR–INO), c/o European Laboratory for Non-Linear Spectroscopy (LENS), Via Nello Carrara 1, Sesto Fiorentino, 50019, Italy

Rocco Duquennoy, Ramin Emadi

Dipartimento di Fisica, Università di Napoli, via Cinthia 21, Fuorigrotta, 80126, Italy

Istituto Nazionele di Ottica (CNR–INO), c/o European Laboratory for Non-Linear Spectroscopy (LENS), Via Nello Carrara 1, Sesto Fiorentino, 50019, Italy

Maria Guadalupe Aparicio

Instituto Nacional de Tecnología Industrial (INTI), Metrología Física, Departamento de Luminotecnia, Laboratorio de Radiometría y Fotometría Básica, Av. Gral. Paz 5445 San Martin, Buenos Aires, Argentina

Maja Colautti

Istituto Nazionele di Ottica (CNR–INO), c/o European Laboratory for Non-Linear Spectroscopy (LENS), Via Nello Carrara 1, Sesto Fiorentino, 50019, Italy

European Laboratory for Non-Linear Spectroscopy (LENS), Via Nello Carrara 1, Sesto Fiorentino, 50019, Italy

Marco López

Physikalisch-Technische Bundesanstalt (PTB), Bundesallee 100, 38116 Braunschweig, Germany

Stefan Kück

Physikalisch-Technische Bundesanstalt (PTB), Bundesallee 100, 38116 Braunschweig, Germany

Laboratory for Emerging Nanometrology, Langer Kamp 6a/b, 38106 Braunschweig, Germany

Costanza Toninelli

Istituto Nazionele di Ottica (CNR–INO), c/o European Laboratory for Non-Linear Spectroscopy (LENS), Via Nello Carrara 1, Sesto Fiorentino, 50019, Italy

European Laboratory for Non-Linear Spectroscopy (LENS), Via Nello Carrara 1, Sesto Fiorentino, 50019, Italy

E-mail: costanza.toninelli@ino.cnr.it

**Abstract**

In the realm of fundamental quantum science and technologies, non-classical states of light, such as single-photon Fock states, are widely studied. However, current standards and metrological procedures are not optimized for low light levels. Progress in this crucial scientific domain depends on innovative metrology approaches, utilizing reliable devices based on quantum effects. We present a new generation of molecule-based single photon sources, combining their integration in a polymeric micro-lens with pulsed excitation schemes, thereby realizing suitable resources in quantum radiometry. Our strategy enhances the efficiency of generated single photon pulses and improves stability, providing a portable source at 784.7 nm that maintains consistent performance even through a cooling and heating cycle. The calibration of a single photon avalanche detector is demonstrated using light sources with different photon statistics, and the advantages of the single-molecule device are discussed. A relative uncertainty on the intrinsic detection efficiency well below 1 % is attained, representing a new benchmark in the field.

## 1. Introduction

The growing interest in quantum technologies has triggered compelling needs in the definition of novel standards and metrology procedures, tailored to capture the physics of quantum systems and quantum states.[1-4] In this context, a prominent role is played by quantum photonics, whereby non-classical light sources in the solid state are combined with the well-established classical technology of miniaturized circuitry for light.[5,6] Within this framework, the development of deterministic single photon sources (SPSs) is a fundamental milestone,[7] as their advent is crucial in many protocols for photon-based quantum simulation,[8,9] computation,[10,11] communication[12,13,14] and metrology.[15,16] While the low-photon-flux regime is challenging with respect to the current standards of radiometry,[17, 18, 19] SPSs potentially offer alternative strategies for delving into regions of very low optical power (pW to fW range)[20, 21,22] and for the calibration of optical components.[23] In particular, they can be used for determining the absolute detection efficiency of Single-Photon Avalanche Detectors (SPADs), which typically suffers from serious artifacts. These are due to several combined effects: the fluctuations in the number of photons per pulse of the light source, the lack of photon-number-resolving power of SPADs, and the presence of a dead time after each detection event. With respect to attenuated laser pulses, SPSs yield direct solutions and countermeasures thanks to their characteristic sub-Poissonian statistics, provided they show sufficiently bright, pure and narrowband emission.

In a broader perspective, quantum emitters in the solid state triggered by pulsed excitation hold promise for the delivery of single photons on demand.[24] A wide range of applications benefit from the variety of platforms, each with specific characteristics. Among others, semiconductor quantum dots are nowadays able to provide high quality sources of single photons,[25,26] entangled-photon pairs[27,28] and squeezed multi-photon states.[29,30]

With respect to metrological applications however, characteristics such as scalability, portability, narrowband emission and flexibility in the fluorescence wavelength become highly valuable, together with fabrication yield and simplicity of operation. For all these reasons, the approach based on organic molecules has become increasingly interesting. Polyaromatic hydrocarbons (PAH), embedded as impurities in solid matrices[31,32,33] and now even in hybrid devices with 2D materials, [34] have proven excellent candidates as single-photon sources for multiple technological applications.[35] On the one hand, they offer a wide palette of emission frequencies from the visible to the near infrared[36] and on the other hand, they can provide bright and stable photon fluxes also at room temperature.[37,38] A pioneering result presented in Ref.[39] e.g., demonstrates intensity fluctuations below the shot-noise limit for light generated by an organic molecule under pulsed operation in ambient conditions. Cooled down to 3 Kelvin, certain PAH molecules show narrowband, highly pure and indistinguishable single-photon emission without the need for optical resonators.[40,41,42,43] Moreover, integration in polymer photonics has been proven to be simple, robust and cost-effective.[44,45] Molecule-based SPSs have already found applications in quantum radiometry,[46] exhibiting up to $1.4 \times 10^6$ photons/s at the fiber-coupled detector. However, due to the continuous wave operation of the source, saturation effects caused by the SPAD dead time were observed, resulting in a count rate dependent detection efficiency $\eta$.

Leveraging the recent advances developed in our group concerning pulsed operation, photon collection,[40] and integration of microlenses,[44] this work presents a new generation of molecular sources and a step forward in the promising path of molecules for quantum radiometry. In particular, the intrinsic efficiency of a single photon detector is measured and compared using different light sources. The analysis of the experimental results, that takes into account the impact of the corresponding photon statistics on the detector dead time, yields consistent results. Finally, the estimated precision in the detector calibration outperforms previous experiments and demonstrates the model- and parameter-free approach enabled by true single photon sources. More in general, together with a similar paper based on quantum-dot SPSs,[47] this work constitutes a milestone towards the uptake of quantum emitters as resources for metrology.

The paper is organized as follows: section 2 contains the description of the setup and of the SPS device used for the metrological application; in section 3 the procedure adopted for the calibration of a silicon SPAD is presented, while the results obtained for a detector in use in our lab are analyzed and discussed in section 4; section 5 reports on the robustness of the proposed design upon warm-up and cool-down cycles, whereas conclusions and outlooks are presented in section 6.

Throughout the paper, with the expression "counts per second" (cps) we refer to the count rate of a SPAD, while "photons per second" (pps) refers to the incoming photon flux at the SPAD, evaluated with the reference analog detector.

## 2. Single photon source and setup description

The calibration of a test detector is performed comparing the count rate of the SPAD with the signal of an analog low-noise reference detector, traceable to the national standard for optical radiant flux, the cryogenic radiometer. Hence, the ideal metrological procedure requires illumination with true single-photon pulses separated by more than the dead time of the SPAD.

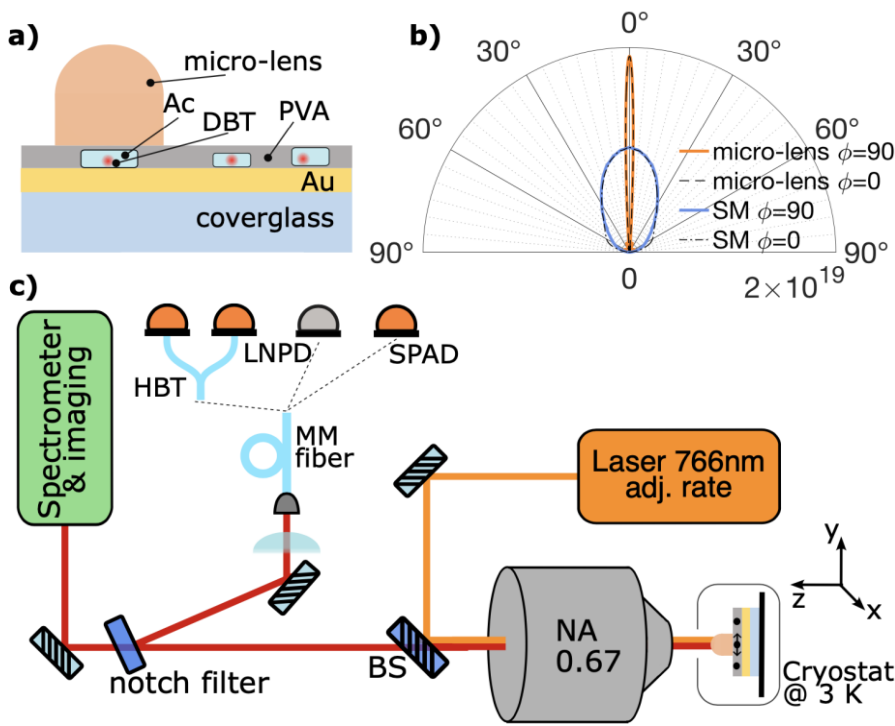

**Figure 1.** a) Artistic view of the SPS device. Sub-micrometric crystals of Anthracene (Ac) doped with Dibenzoterrylene (DBT) molecules are dispersed on a gold surface (Au) acting as a mirror and are covered with a few hundred-nanometer-thick layer of Poly-Vinyl Alcohol (PVA). The polymeric nano-optic element (micro-lens) fabricated on pre-selected crystals is drawn in orange. b) Angular distribution of the power out-flow for the molecule-based SPS device before and after the integration of the micro-lens. c) Sketch of the experimental setup. By means of a notch filter, the molecule emission collected in an epifluorescence-microscope configuration is separated in two parts containing the zero-phonon line (ZPL) and the phonon sideband (PSB). Both components can then be analyzed with a spectrometer. The ZPL component can also enter the fibered circuit for probing the purity of the photon wave packet (HBT) or for implementing the calibration of a SPAD against a low-noise reference analog detector (LNPD).

The design of the SPS exploited in this paper is sketched in **Figure 1a**. Sub-micrometric crystals (NCs) of Anthracene doped with Dibenzoterrylene molecules (DBT:Ac) are dispersed on a gold mirror and protected by a few-hundred-nm-thick layer of poly-vinyl alcohol (PVA). Weierstrass-like polymeric micro-lenses are then fabricated with the direct-laser-writing technique on top of NCs selected after preliminary characterization of the fluorescence emitted at room temperature.

Details on NCs synthesis and characterization can be found in Ref. [38], while the NCs-on-mirror design and a preliminary integration with polymeric micro-lenses are discussed in Ref.[46] and Ref.[44], respectively. The morphological properties of NCs guarantee that such a simple preparation is able to provide emitters with the optical dipole aligned in the horizontal plane and at a distance of around 100 nm from the gold mirror, maximizing the emission directionality within a small angle around the polar axis.[48] The introduction of an integrated micro-lens brings three additional advantages: i) further enhancement of the directionality; ii) reduction of the losses into the guided modes at the metal-PVA interface; iii) creation of an additional protective structure, which makes the SPS robust against multiple temperature cycles, ranging from room temperature down to cryogenic levels. An estimation of the expected emission pattern, according to finite element simulations, is depicted in Figure 1b.

With respect to the process discussed in Ref. [44], the following differences are noteworthy. The nanocrystals have much lower DBT concentration in order to allow for off-resonant pumping of single molecule emission. In this way, spectral selection of the ZPL is possible, allowing for narrow band operation and also for their use in quantum protocols based on two-photon interference.

The experimental setup employed for the characterization of the SPS and for the calibration of the SPAD is sketched in Figure 1c. Single-molecule fluorescence is excited and collected in an epi-fluorescence microscope through a 0.67-N.A. objective lens (SigmaKoki PAL-50-NIR-HR-LC07, transmission at 785 nm = 0.7), using a pulsed laser with adjustable repetition rate and 50-ps long pulses, which can be operated in CW as well (Picoquant LDH-D-FA-765L, linewidth of about 0.3 nm at 759.9 nm). A long-pass filter (Semrock LP02–785RE) is employed for rejection of the back-scattered pump light. In order to maximize the collection of the bright component at around 784.5 nm (the so called 00-Zero-Phonon-Line (ZPL)), a beam-sampler with 8 % reflection and more than 90 % transmission with almost flat response (Thorlabs

BSF20-B) is used as semi-reflective element separating input and output from the objective lens. Moreover, the ZPL is filtered out from the phonon sideband (PSB) exploiting a 0.4 nm-wide reflective notch filter (OptiGrate BNF-785-OD4–12.5M) and then coupled to a 0.22 N.A., 105 $\mu$m core diameter, multi-mode fiber (Thorlabs FG105LCA) via a 100-mm focal length plano-convex lens, exhibiting coupling efficiency beyond 0.9. Both ZPL and PSB can be imaged or spectrally analyzed with an Andor-Solis system as well (SR-303i-A, camera iXon3). The fiber-coupled photon flux can be delivered either to a fiber-based Hanbury-Brown and Twiss (HBT) setup for purity estimation, or to a FC/SC to E-2000 adaptor sleeve for performing the calibration task. The first part is constituted by a non-polarizing multi-mode beam-splitter (Thorlabs TM105R5F2B), a couple of SPADs (Excelitas SPCM-NIR-14), and a time tagger time domain module (QuTools quTAG standard). The rest of the setup is further described in section 3. Once a promising micro-lensed NC is selected by a qualitative analysis in widefield imaging, confocal illumination (and detection) is adopted for the optical characterization of the source. The use of isolated NCs enables single emitter addressing without specific spatial filtering beyond confocal microscopy. **Figure 2** depicts the data collected for the device employed in the following for the SPAD calibration.

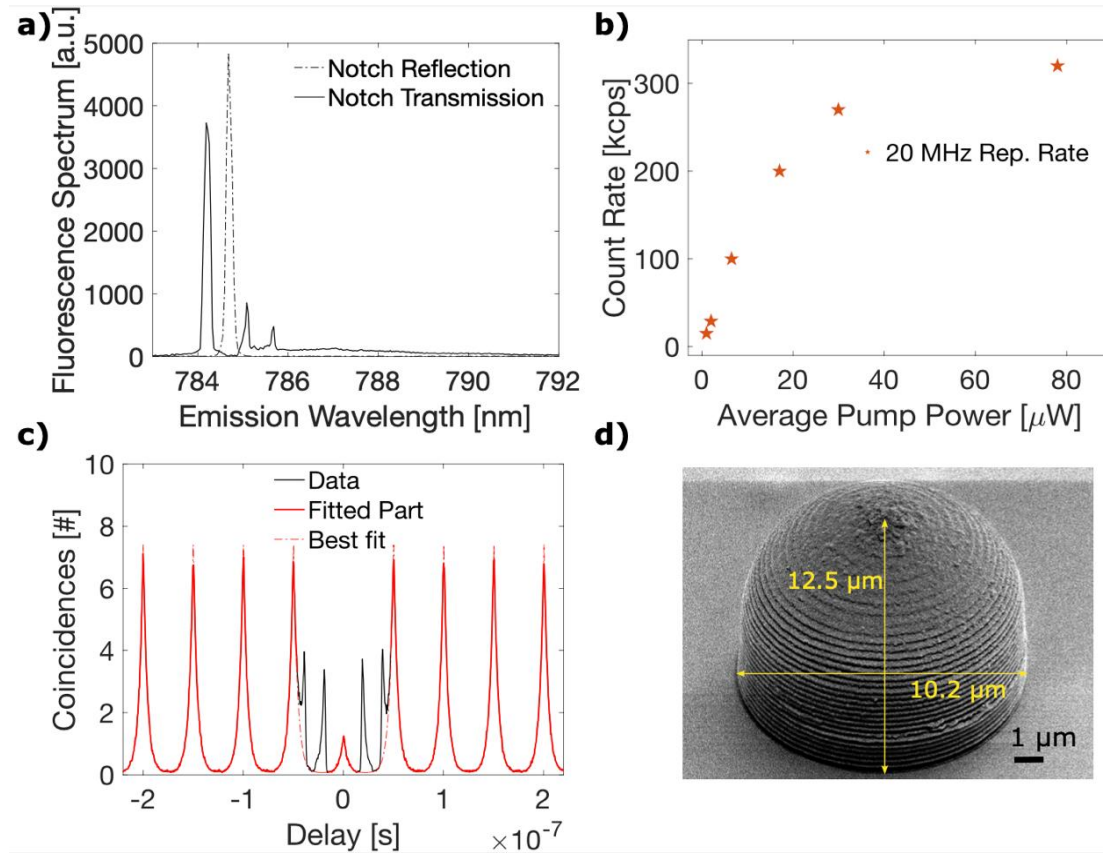

**Figure 2.** Characterization of the device employed for the calibration. All the measurements reported in panel a)-c) have been taken while operating the device at 20-MHz repetition rate. a) Fluorescence spectrum: solid and dashed lines correspond to the components as separated by the notch filter, transmitted and reflected, respectively. Each peak belongs to a different molecule, the spectral selection is tight enough to guarantee collection of photons from a single emitter (at 784.7 nm). b) Photon flux detected with the SPAD as a function of the laser pump power. c) Normalized histogram of the inter-photon arrival times for 60 μW average pump power (black solid line). The additional peaks at around 20 ns and 40 ns from the central one are due to afterpulsing and afterglow events from the detectors, the latter made relevant by the back-reflections from the flat facets of the multi-mode fiber components used in the HBT setup. The red line is a fit to the part of the data not affected by the afterglow (marked in blue) with the expression defined in Ref. [40], holding $g^{(2)}[0]=0.17+/-0.02$ . d) Scanning electron microscope image of the device.

The emission spectrum is reported in Figure 2a, where one can clearly distinguish the presence of two bright lines separated by about 0.5 nm, which are associated with two different emitters in the same NC, plus a couple of weakly addressed nearby emitters. Thanks to the combined effect of the tight spectral selection applied by the notch filter (FWHM = 0.4 nm) and the spatial selection through the fiber coupling, the ZPL of a single molecule can be isolated in the reflection from the notch filter (dotted line, centered at 784.7 nm). The spectral width of around 0.2 nm (shown in the figure), is narrow enough for accurate radiant flux measurements at a well-defined wavelength, and is actually given by the spectrometer resolution. In fact, typical ZPL linewidth for DBT:AC systems at 3 Kelvin are well below 500 MHz.[38,44]

The saturation curve (Figure 2b) shows that for an average pump power above 50 $\mu$W and a repetition rate of 20 MHz, the device can provide a photon flux of the order of 300 kcps at the detector. According to the calibration reported in section 3, this corresponds to a power beyond 100 fW, which is high enough to be measured with a low-noise analog reference detector (LNPD). Finally, Fig. 2c represents the emitted intensity autocorrelation histogram for single-photon purity corresponding to the condition of maximum photon flux, and 20-MHz repetition rate of the pump laser. A clear anti-bunching feature characterized by $g^{(2)}[0]=0.17+/-0.02$ confirms the low multi-photon probability of the generated stream of pulses. The $g^{(2)}[0]$ is determined as the ratio between the amplitude of the peak at zero delay and the other peaks, both evaluated as best fit to the data (see Ref.[40] for details on the fitting procedure and function). Figure 2d shows a scanning electron microscope image of the microlens structure integrating the NC. The 45 degrees tilt of the sample substrate against the electronic beam allows estimating the characteristic dimensions of the nanofabricated device, highlighted in the figure. The target structure configuration is based on a modified Weierstrass micro-len,[44] consisting of a hemispherical dome and a cylindrical basis with a total height $h = (1 + 1/n)a$, where $a$ is the cylinder radius and $n$ is the refractive index. Based on Figure 1b, a maximum collection efficiency of about 40 % is expected, but upon optimization of the fabrication process we can aim for an improvement of about a factor of two.

## 3. Detector calibration

The single-photon flux characterized above is high enough to be measured with an LNPD. The Optics division at PTB hence provided an LNPD traced to the national standard for optical radiant flux and with working range down to few tens of femtowatt (Femto, FWPR-20-S). Such a device allows for a direct calibration of a SPAD by simply relating the click rate $N_{click}$ to the actual photon flux $\phi$. As a demonstrative case, we report here the calibration of one of the detectors in use in the HBT setup (Excelitas SPCM-NIR-14, S.N. 37207).

In order to have a proper calibration, however, correction factors accounting for not ideal behavior of the devices need to be included. Indeed, the real efficiency can be overestimated because of the presence of additional fake counts in $N_{click}$ due to the e afterpulsing phenomenon in the SPAD. Moreover, the impact of residual multi-photon events that result in single clicks should be considered, in order to avoid underestimation. According to Refs. [49,47], these corrections can be introduced in the expression for the evaluation of the quantum efficiency $\eta$ of the SPAD as follows:

$$\eta = \frac{N_{click}}{\phi} \frac{(1-p_A)}{(1-\varepsilon)} \qquad (1)$$

where $p_A$ is the afterpulsing probability for each click event, and $\varepsilon$ represents the probability of having more than one photon per pulse. The latter is estimated from the $g^{(2)}[0]$ and the mean photon number per pulse $\phi/R$ as $\varepsilon = \frac{1}{2}g^{(2)}[0]\frac{\phi}{R}$, with R being the repetition rate of the source trigger. In our case, based on the source characterization performed in the previous section, we can estimate a multi-photon-induced deviation of $\eta$ of the order of $10^{-3}$, that will result to be negligible in the following discussion. Another source of systematic error for $N_{click}$ is given by the dark count rate $N_{DC}$ of the SPAD and its interplay with the dead time D, defined as the time window following a detection event in which the detection of photons is inhibited. However, the weight of the latter contribution is determined by the product $N_{DC}*D$,[50] which is of the order of $10^{-5}$ for our detector ($N_{DC} < 10^3 s^{-1}$ and D ~ 2 $10^{-8}$ s), and will be neglected in the following.

The photon flux $\phi$ is determined by means of the LNPD signal $U_{LNPD}$ as

$$\phi = \frac{\lambda}{h\,c} P = \frac{\lambda}{h\,c} \frac{U_{LNPD}}{s_{LNPD}} \qquad (2)$$

where $\frac{\lambda}{h\,c}$ is the energy of a photon (h - Planck constant, c - speed of light, lambda - emission wavelength: 784.7 ± 0.1 nm), and P the optical radiant flux estimated by means of the LNPD as $U_{LNPD}/s_{LNPD}$ (corresponding to the signal at the output voltage of the LNPD in Volts and the spectral responsivity of the LNPD in Volt per Watt, respectively). Traceability to the primary standard for optical power, the cryogenic radiometer, is achieved by a calibration of $s_{LNPD}$. This calibration has been conducted separately according to the double attenuator technique,[51] reading an ultimate value $s_{LNPD} = (0.5562 \pm 0.0019)\,10^{12}\,V/W$ at the wavelength of interest, and with not detectable non-linearity within the range of power relevant for this work. Also, the LNPD exhibits a non-zero output signal $U_0$ corresponding to $\phi = 0$, which needs to be subtracted for a correct estimation of $\eta$.

In order to determine the optimal procedure for the calibration, we have first probed the statistical properties of the produced photon stream. The inset of **Figure 3a** displays a typical count trace from which the flux stability has been evaluated in terms of an Allan deviation (Figure 3a, main panel). As expected due to the negligible blinking of the system,[52,38] the photon flux appears very stable, reaching a relative uncertainty of less than $10^{-3}$ for integration times of the order of one minute. For longer times, we detect an almost linear drift of the average count rate, which a successive investigation has attributed to the drift of the confocal spot due to a motorized mirror mount. Moreover, the LNPD zero-flux signal $U_0$ shows a strong dependence on temperature, resulting in a relevant drift of the background mean value.

In order to optimize accuracy and precision, the measurements were hence performed as follows: the photon stream is sent to the SPAD and then, immediately after, to the calibrated LNPD; the sequence is terminated by a second measurement with the SPAD. Each step lasts for around one minute, as suggested from the Allan deviation plot. For both detectors, reference level traces are acquired while the signal is delivered to the other device. This procedure is useful to control and mitigate errors coming either from a drift of the dark current of the LNPD or from a change in the photon flux. Particular attention is also paid to minimize any circumstance which can modify the flux during the whole operation: each detector has its own multi-mode fiber patch cable terminated with E-2000 connector, and the only element manipulated during the calibration sequence is the FC/PC to E-2000 adaptor sleeve, which in turns guarantees high reproducibility of the fiber-to-fiber coupling.

Considering Equation 1 and 2, and the procedure described above, the expression used for the calibration becomes:

$$\eta = \frac{h\,c}{\lambda}\,s_{LNPD}\,\frac{(N^{(1)}_{click}+N^{(2)}_{click})/2 - N_{DC}}{U-(U^{(1)}_0+U^{(2)}_0)/2}(1-p_A) \qquad (3)$$

The complete procedure has been repeated for each data point in Figure 3b, where the quantum efficiency of the same device is reported for different photon fluxes and photon source types. In particular, we have compared the SPS presented in the previous sections of the paper (pSPS), a similar source operated under continuous wave pumping (cwSPS), and attenuated pulses from the laser used to excite the SPS (pL).

At first glance, it is possible to recognize the monotonic behavior of the detection efficiency, which reaches its maximum ($\eta \sim 0.66$) for low photon flux, where the probability of photon arrival during the dead time goes to zero. Moreover, we observe that the value obtained with the pSPS is higher than that obtained from the pL and the cwSPS.

The uncertainty attributed to each data point is carefully estimated, as discussed in the Supporting Information, and the results are detailed in **Table S1**. The uncertainty budget for the case of the pSPS operated at $R = 60$ MHz is presented as a typical example. We find that the overall uncertainty is dominated by the terms ascribed to the LNPD (fluctuations in the dark signal and uncertainty in the determination of $s_{LNPD}$), which confirms the high stability and intensity of the photon flux provided by our device for application in quantum radiometry.

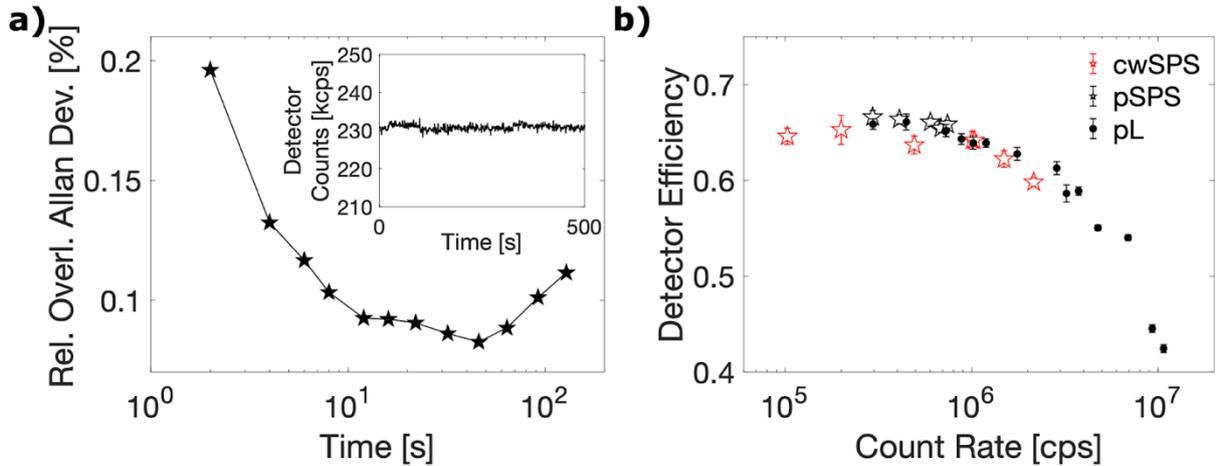

**Figure 3.** a) Statistical properties of the photon stream produced by the pSPS at $R = 20$ MHz and 36 μW average pump power in terms of Allan deviation of $N_{click}$. Minimum uncertainty (below 0.1%) is obtained for an integration time between 10 s and 50 s. The inset shows the $N_{click}$ trace from which the flux stability has been evaluated. b) Detector efficiency $\eta$ of the SPAD obtained with photon fluxes of various intensities and origins (source types). The measurements taken with pSPS read the highest values for $\eta$, and we report a general monotonic descending trend for higher fluxes, given by the growth of the probability of receiving photons during the dead time following each detection event.

## 4. Discussion

In order to extract the intrinsic detector efficiency $\eta_0$ from the presented experimental results, a model of the SPAD response accounting for the impact of the dead time as a function of the different photon statistics is required. Following Ref. [50], if we define the click probability per trigger event as q, then $N_{click}$ is determined by q, D (detector dead time) and R (repetition rate) following the equation

$$R * q = N_{click} + Int[R * D] * q * N_{click}$$

Indeed, for any detection event we expect on average a number of lost events equal to $Int[R * D] * q$, where $Int[]$ stands for the lower integer part, and $Int[R * D]$ corresponds to the (deterministic) number of pulses falling in the dead time period after a detection event. From the equation above, we obtain a general expression for $\eta$:

$$\eta = \frac{N_{click}}{\phi} = \frac{R}{\phi} \frac{q}{1+Int[R*D]q} \qquad (4)$$

According to the origin of the photon stream used for the calibration of the detector, the terms in Equation 4 assume different shapes. The specific equations valid for the different illumination sources are summarized in **Table 1**.

**Table 1.** Model functions according to the light source type, based on Ref. [50]

| Parameter / Source Type | $\phi$ | R*q | notes | expression |
|---|---|---|---|---|
| pulsed Laser (pL) | $R * \mu$ | $R(1 - e^{-\mu*\eta_0})$ | $\mu$: Poissonian mean photon # | $\eta = \frac{1}{\mu} \frac{(1 - e^{-\mu*\eta_0})}{1 + Int[R * D](1 - e}$ |
| pulsed SPS (pSPS) | $R * \eta_s$ | $R * \eta_s * \eta_0$ | $\eta_s$: SPS efficiency | $\eta = \eta_0 \frac{1}{1 + Int[R * D]\eta_s * \eta}$ |
| cw SPS (cwSPS) | $r$ | $r * \eta_0$ | r: photon rate at detector; $Int[R * D]q \rightarrow r * D * \eta_0$ | $\eta = \eta_0 \frac{1}{1 + r * D * \eta_0}$ |

where $\eta_0$ represents the intrinsic efficiency of the detector. A fundamental aspect which becomes evident from the modeling above, is that the analysis for the evaluation of $\eta_0$ depends on the type of source, and hence we have to address the different cases separately. In particular, a step-like trend is expected for a pSPS as a function of the repetition rate, regardless of the source efficiency $\eta_s$, as the interphoton time interval is fixed. Instead, a monotonous descending trend characterizes instead CW sources, as an increase of the flux $r$ is obtained by pumping harder. A calibration based on pL shows a combination of both trends. These behaviors are a direct consequence of the probability of photon arrival during the dead times.

**Figure 4** shows the results of the analysis based on the model presented above, with the estimation of $\eta_0$ as best fit to the data for the calibration with the pSPS and the pL (panel a, c and d) and with the cwSPS (panel b). Interestingly, even though a proper modeling gives access to $\eta_0$ for each considered source, exploiting an SPS under pulsed operation one can benefit from a direct estimation, without assumption on the model or on the value of the other quantities. Indeed, the "first" step of the $\eta$ vs $R$ curve, where the separation between successive single-photon pulses is longer than the dead time of the SPAD (meaning that $Int[R*D]$ vanishes), exactly yields $\eta_0$, regardless of any other parameter. This is in contrast to what can be obtained with pL, for which the inherent photon statistics determines instead a condition $\eta < \eta_0$, for any value of the mean photon number per pulse $\mu$. In the analysis we have omitted the data obtained for R = 50 MHz, as this value is very close to $1/D$ and hence lies in an intermediate range with no well determined behavior.

The calibrations performed with the pSPS and the cwSPS show a good agreement with the model and give consistent results. Concerning the dataset related to the pL measurements, only a partial agreement is obtained, and the results of the fit for the different values of $\mu$ are not fully consistent. In particular, the $\eta_0$ obtained from measurements with low $\mu$ converges to the value found with the molecular source. A deviation is instead observed especially for a high average number of photons. This can be due to the limited applicability of the model, that was developed and tested for InGaAs SPADs, characterized by different quenching circuits.

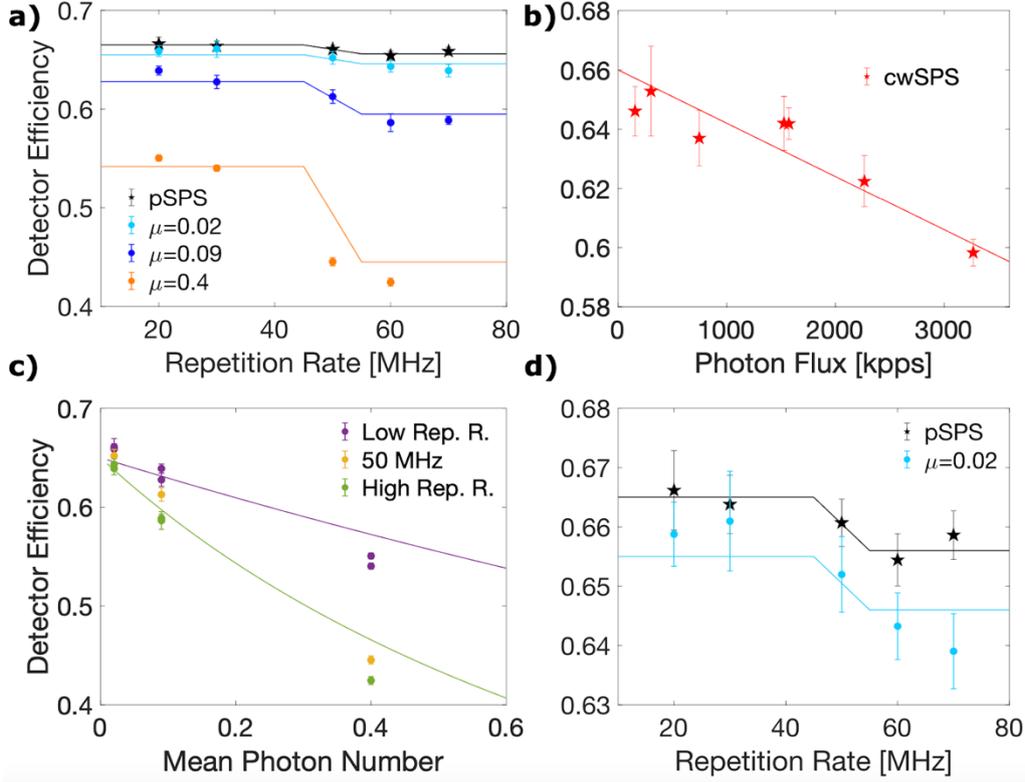

**Figure 4.** Comparing detection efficiency measurements. Experimental data (scatters) and best fit to the data (solid lines) are presented for SPSs (stars) and pL (bullets). a) $\eta$ as a function of the repetition rate R for pSPS and pL. b) $\eta$ as a function of the average photon flux $r$ for cwSPS. In this case $\eta_0$ is determined as the intercept of a linear fit, corresponding to an approximation of the model for low $r$. c) $\eta$ as a function of $\mu$ for pL, setting $\eta_0$ equal to the weighted average of the fitting outputs in a). d) Zoom of the data of panel a) to compare pSPS and pL for low $\mu$: both the fit and the data are consistent with the expected reduction of $\eta$ by a factor ~ 0.005.

**Table 2** summarizes the findings. Notably, the calibration reported for the pSPS yields a relative uncertainty of 0.3 % ($\eta_0 = 0.0665 \pm 0.0002$), and outperforms the results obtained with the same technique exploiting a quantum dot-based device,[47] despite the higher photon flux achieved in that work. This is probably due to the photo-stability of the molecular sources, combined with the measurement protocol chosen to leverage on it. A consistent result, even though with lower precision, is obtained for an SPS operated in CW mode. Concerning the calibration with the pL instead, the measurement at low mean photon number ($\mu = 0.02$) reads $\eta_0 = 0.659 \pm 0.03$, while a weighted average of the results obtained for the different $\mu$ yields a final estimation of $\eta_0 = 0.65$. This value has been used for the comparison between experimental findings and model in panel c, where $\eta$ is presented as a function of $\mu$, showing only partial agreement. The fitting operation is performed with Wolfram Mathematica (NonlinearModelFit function), yielding also the uncertainty estimation for each data point. Finally, panel d represents a zoom of the data in panel a, highlighting the difference in the measured efficiency between the laser and our true SPS due to the photon statistics. As expected from a simple Taylor expansion of the model, the gap results of the order of $\mu \eta_0^2/2 \sim 0.005$.

**Table 2.** Intrinsic SPAD Detection efficiency $\eta_0$ obtained from the calibration with different sources

| source type | pSPS | cwSPS | pL, µ =0.02 | pL, µ=0.09 | pL, µ=0.4 |
|---|---|---|---|---|---|
| $\eta_0$ | 0.665±0.002 | 0.660±0.007 | 0.659±0.003 | 0.647±0.005 | 0.61±0.01 |
| relative uncertainty | 0.3% | 1% | 0.5% | 0.8% | 1.6% |

## 5. Sample stability across multiple warm-up cool-down cycles

Quantum light sources based on organic compounds are generally considered thermally unstable in literature, as bare samples undergo detrimental aging on a short time scale and are soluble in generic organic solvents.[53,54] In this respect, devices based on PVA-coated NCs have shown a remarkable mitigation of the issue, mostly related to the sublimation of the host matrix. The design presented in this work represents a further step forward in the direction of durable organic SPSs. In order to determine the robustness and reproducibility of the optical performances, we investigate the behavior of a single device in a series of cooldowns. The device discussed in the previous sections is analyzed a second time at low temperature, after three days of rest at room temperature and passive vacuum in the cryostat. Under the very same excitation and collection conditions, we have found a negligible variation of flux and purity of the single-photon stream. In detail, we report a slightly higher maximum photon flux of 340 kcps for a slightly less average pump power of 58 µW (at 20 MHz repetition rate and same excitation wavelength 765.9 nm). Probing the second-order autocorrelation function we reads a $g^{(2)}[0]= 0.21 \pm 0.03$ as best fit to the data, corresponding to a multi-photon probability of ~ 0.0023 (against the 0.0019 obtained during the first cooldown, see Fig. 5a for comparison). Moreover, probing the emission spectrum of the device, we have detected a shift of the ZPL peak by around 0.2 nm to the blue (Fig. 5b), which is smaller than the selection band of the notch filter and allows the operation of the source without any adjustment in the setup.

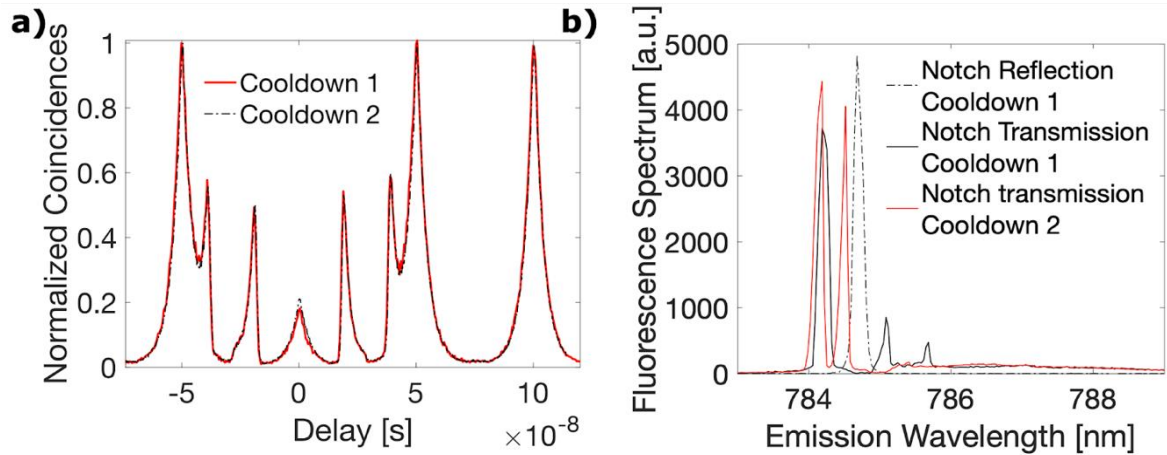

**Figure 5** Durability of the device: comparison of the photon flux characteristics in different cooldowns. a) Normalized histogram of the inter-photon arrival times for 60 µW average pump power; best fits to the data read $g^{(2)}[0]=0.17\pm0.02$ and $g^{(2)}[0]=0.21\pm0.03$, respectively. b) Emission spectrum: the measurement performed in the second cooldown (red line) is obtained after shifting the notch reflection window to the red by around 0.5 nm for simplicity. The lines of the two bright molecules are shifted to the blue by around 0.2 nm.

## 6. Conclusions and Outlook

In this paper, the calibration of a Single Photon Avalanche Detector by means of a molecule-based Single-Photon Source under pulsed operation is achieved for the first time. Thanks to the statistical properties of the produced photon flux, the attained precision (below 1%) outperforms the results in literature for other sources, even in the case of higher maximum photon flux. The device shows enhanced durability, yielding consistent performance after multiple cooling and heating cycles. These results are obtained thanks to the nano-fabrication of a polymeric micro-lens on preselected molecules via two-photon polymerization direct laser writing technique, providing a protective cap but also an enhancement in the collection efficiency. In essence, we demonstrate that molecule-based single-photon sources under pulsed operation allow for a direct estimation of detectors' efficiency, without any assumption on other parameters including the source brightness, making this work of high metrological importance. In principle, the performance is expected to improve further as the average photon number per pulse reaches values close to unity, leveraging intensity fluctuations below the shot-noise limit. In this respect, we expect an improved efficiency at detector beyond 10%, upon optimization of the micro-lens design and spectral selection of the molecule population with best quantum yield.[55] Finally, considering the implemented pumping schemes that allows for collecting light from the Fourier-limited ZPL, the devised single photon source could have impact not only on quantum radiometry, but also in optical quantum communication or simulation protocols based on indistinguishable photons on-demand.

**Supporting Information**

Supporting Information is available and contains the Table that summarizes all the contributions to the uncertainty on the detection efficiency estimation and the evaluation of the afterpulsing probability.

**Acknowledgments**

This work is financed by the EMPIR programme (Project No. 20FUN05, SEQUME), cofinanced by the Participating States and by the European Union's Horizon 2020 Research and Innovation Programme. The research has been co-funded by the European Union - NextGeneration EU, "Integrated infrastructure initiative in Photonic and Quantum Sciences" - I-PHOQS [IR0000016, ID D2B8D520, CUP B53C22001750006]. It is also co-funded by the European Union (ERC, QUINTESSEnCE, 101088394). Views and opinions expressed are however those of the author(s) only and do not necessarily reflect those of the European Union or the European Research Council. Neither the European Union nor the granting authority can be held responsible for them. The authors wish to acknowledge Elisa Riccardi and Miriam Vitiello from Istituto Nanoscienze-CNR for their support with SEM measurements and Felix Binkowski and Sven Burger for their help with JCM suite for numerical simulations.

**References**

[1] O. van Deventer, N. Spethmann, M. Loeffler, M. Amoretti, R. van den Brink, N. Bruno, P. Comi, N. Farrugia, M. Gramegna, A. Jenet, B. Kassenberg, W. Kozlowski, T. Länger, T. Lindstrom, V. Martin, N. Neumann, H. Papadopoulos, S. Pascazio, M. Peev, R. Pitwon, M. A. Rol, P. Traina, P. Venderbosch, F. K. Wilhelm-Mauch, *EPJ Quantum Technol*. **2022**, *9,* 33.
[2] A. M. Bhargav, R. K. Rakshit, S. Das, M. Singh, *Adv. Quantum Technol.* **2021**, *4*, 10.
[3] M. Barbieri, *PRX Quantum* **2022**, *3*, 010202.
[4] Quantum Flagship Strategic Research and Industry Agenda, https://qt.eu/media/pdf/Quantum-Flagship_SRIA_2022_0.pdf
[5] J. L. O'Brien, A. Furusawa, J. Vuckovic, *Nat. Photon.* **2009**, *3*, 687.
[6] J. Wang, F. Sciarrino, A. Laing, M. G. Thompson, *Nat. Photon.* **2020**, *14(5)*, 273-284.
[7] N. Sangouard, H. Zbinden, *J. Mod. Opt.* **2012**, *59*, 1458.
[8] A. Aspuru-Guzik, P. Walther, *Nat. Phys.* **2012***, 8(4)*, 285-291.

[9] C. Sparrow, E. Martín-López, N. Maraviglia, A. Neville, C. Harrold, J. Carolan, Y. N. Joglekar, T. Hashimoto, N. Matsuda, J. L. O'Brien, D. P. Tew, A. Laing, *Nature* **2018**, *557(7707)*, 660-667.

[10] T. D. Ladd, F. Jelezko, R. Laflamme, Y. Nakamura, C. Monroe, J. L. O'Brien, *Nature* **2010**, *464*, 45.

[11] J. B. Spring, B. J. Metcalf, P. C. Humphreys, W. S. Kolthammer, X.-M. Jin, M. Barbieri, A. Datta, N. Thomas-Peter, N. K. Langford, D. Kundys, J. C. Gates, B. J. Smith, P. G. R. Smith, I. A. Walmsley, *Science* **2013**, *339*, 798.

[12] V. Scarani, H. Bechmann-Pasquinucci, N. J. Cerf, M. Dušek, N. Lütkenhaus, M. Peev *Rev. Mod. Phys.* **2009**, *81(3)*, 1301.

[13] W. Luo, L. Cao, Y. Shi, L. Wan, H. Zhang, S. Li, G. Chen, Y. Li, S. Li, Y. Wang, S. Sun, M. F. Karim, H. Cai, L. C. Kwek, A. Q. Liu, *Light Sci. Appl.* **2023**, *12(1)*, 175.

[14] H. J. Kimble, *Nature* **2008,** *453*, 1023.

[15] C. J. Chunnilall, I. P. Degiovanni, S. Kück, I. Müller, A. G. Sinclair, *Opt. Eng.* **2014***, 53(8)*, 081910 (August 2014)

[16] C. Couteau, S. Barz, T. Durt, T. Gerrits, J. Huwer, R. Prevedel, J. Rarity, A. Shields, G. Weihs, *Nat. Rev. Phys.* **2023**, *5,* 354–363.

[17] N. P. Fox, *Metrologia* **1995**, *32*, 535.

[18] T. Dönsberg, M. Sildoja, F. Manoocheri, M. Merimaa, L. Petroff, E. Ikonen, *Metrologia* **2014**, *51,* 197.

[19] J. C. Zwinkels, E. Ikonen, N. P. Fox, G. Ulm, M. L. Rastello, *Metrologia* **2010***, 47,* R15-R32

[20] W. Schmunk, M. Rodenberger, S. Peters, H. Hofer, S. Kück, *J. Mod. Opt.* **2011**, *58:14*, 1252-1259.

[21] B. Rodiek, M. López, H. Hofer, G. Porrovecchio, M. Smid, X.-L. Chu, S. Götzinger, V. Sandoghdar, S. Lindner, C. Becher, S. Kück, *Optica* **2017**, *4*, 71-76.

[22] S. Kück, *Meas.: Sens.* **2021**,*18,* 100219

[23] F. Li, T. Li, M. O. Scully, G. S. Agarwal, *Phys. Rev. Applied* **2021**, *15*, 044030

[24] B. Lounis, M. Orrit, *Rep. Prog. Phys.* **2005,** *68*, 1129–1179

[25] H. Wang, Y. M. He, T. H. Chung, H. Hu, Y. Yu, S. Chen, X. Ding, M. C. Chen, J. Qin, X. Yang, R.-Z. Liu, Z.-C. Duan, J.-P. Li, S. Gerhardt, K. Winkler, J. Jurkat, L.-J. Wang, N. Gregersen, Y.-H. Huom Q. Dai, S. Yu, S. Höfling, C.-Y. Lu, J.-W. Pan, *Nat. Photon*. **2019**, *13*, 770-775.

[26] N. Tomm, A. Javadi, N. O. Antoniadis, D. Najer, M. C. Löbl, A. R. Korsch, R. Schott, S. R. Valentin, A. D. Wieck, A. Ludwig, R. J. Warburton, *Nat. Nanotech.* **2021**, *16*, 399-403.

[27] T. J. Steiner, J. E. Castro, L. Chang, Q. Dang, W. Xie, J. Norman, J. E. Bowers, G. Moody, *PRX Quantum* **2021**, *2(1)*, 010337.

[28] D. Huber, M. Reindl, Y. Huo, H. Huang, J. S. Wildmann, O. G. Schmidt, A. Rastelli, R. Trotta, *Nat. Commun.* **2017**, *8(1)*, 15506.

[29] G. Moody, L. Chang, T. J. Steiner, J. E. Bowers, *AVS Quantum Sci*. **2020**, *2(4)*, 041702

[30] C. Cui, C. N. Gagatsos, S. Guha, L. Fan, *Phys. Rev. Res.,* **2021**, *3(1)*, 013199.

[31] B. Kozankiewicz, M. Orrit, M., *Chem. Soc. Rev.* **2014**, *43*(4), 1029-1043.

[32] M. Orrit, T. Ha, V. Sandoghdar, *Chem. Soc. Rev.* **2014**, *43*(4), 973-976.

[33] R. C. Schofield, P. Burdekin, A. Fasoulakis, L. Devanz, D. P. Bogusz, R. A. Hoggarth, K. Major, A. S. Clark, *Chem. Phys. Chem.* **2022,** *23(4)*, e202100809.

[34] R. Smit, A. Tebyani, J. Hameury, S. J. van der Molen, M. Orrit, *Nat. Commun.* **2023***, 14,* 7960.

[35] C. Toninelli, I. Gerhardt, A. S. Clark, A. Reserbat-Plantey, S. Götzinger, Z. Ristanović, M. Colautti, P. Lombardi, K. D. Major, I. Deperasinska, W. H. Pernice, F. H. Koppens, B. Kozankiewicz, A. Gourdon, V. Sandoghdar, M. Orrit, *Nat. Mat.* **2021**, *20(12)*, 1615-1628.

[36] P. Siyushev, G. Stein, J. Wrachtrup, I. Gerhardt, *Nature* **2014**, *509(7498)*, 66-70.

[37] C. Toninelli, K. Early, J. Bremi, A. Renn, S. Götzinger, V. Sandoghdar, *Opt. Express* **2010,** *18 (7),* 6577-6582.

[38] S. Pazzagli, P. Lombardi, D. Martella, M. Colautti, B. Tiribilli, F. S. Cataliotti, C. Toninelli, *ACS Nano* **2018,** *12(5)*, 4295-4303.

[39] X.-L. Chu, S. Götzinger, V. Sandoghdar, *Nat. Photon.* **2017**, *11*, 58-62.

[40] P. Lombardi, M. Colautti, R. Duquennoy, G. Murtaza, P. Majumder, C. Toninelli, *Appl. Phys. Lett.* **2021**, 118, 204002.

[41] M. Rezai, J. Wrachtrup, I. Gerhardt, *Phys Rev X* **2018**, *8(3)*, 031026

[42] R. C. Schofield, C. Clear, R. A. Hoggarth, K. D. Major, D. P. S. McCutcheon, A. S. Clark, *Phys. Rev. Res.* **2022**, *4 (1),* 013037

[43] R. Duquennoy, M. Colautti, R. Emadi, P. Majumder, P. Lombardi, C. Toninelli, *Optica* **2023,** *9(7)*, 731-737

[44] M. Colautti, P. Lombardi, M. Trapuzzano, F. S. Piccioli, S. Pazzagli, B. Tiribilli, S. Nocentini, F. S. Cataliotti, D. S. Wiersma, C. Toninelli, *Adv. Quantum Technol.* **2020**, *3(7)*, 2000004

[45] D. Rattenbacher, A. Shkarin, J. Renger, T. Utikal, S. Götzinger, V. Sandoghdar, *Optica* **2023**, *10(12)*, 1595-1601

[46] P. Lombardi, M. Trapuzzano, M. Colautti, G. Margheri, I. P. Degiovanni, M. López, S. Kück, C. Toninelli, *Adv. Quantum Technol.* **2020**, *3(2)*, 1900083

[47] H. Georgieva, M. López, H. Hofer, N. Kanold, A. Kaganskiy, S. Rodt, S. Reitzenstein, S. Kück, *Opt. Express* **2021**, *29(15)*, 23500-23507

[48] S. Checcucci, P. Lombardi, S. Rizvi, F. Sgrignuoli, N. Gruhler, F. B. C. Dieleman, F. S. Cataliotti, W. H. P. Pernice, M. Agio, C. Toninelli, *Light Sci. Appl.* **2017**, *6*, e16245

[49] H. Georgieva, M. López, H. Hofer, J. Christinck, B. Rodiek, P. Schnauber, A. Kaganskiy, T. Heindel, S. Rodt, S. Reitzenstein, S. Kück, *Metrologia* **2020**, *57*, 055001

[50] H. Georgieva, A. Meda, S. M. F. Raupach, H. Hofer, M. Gramegna, I. P. Degiovanni, M. Genovese, M. López, S. Kück, *Appl. Phys. Lett.* **2021**, *118*, 174002

[51] M. López, H. Hofer, S. Kück, *J. Mod. Opt.* **2015**, *62*, 1732–1738.

[52] A. A. Nicolet, C. Hofmann, M. A. Kol'chenko, B. Kozankiewicz, M. Orrit, *Chem. Phys. Chem.* **2007**, *8*(8), 1215-1220.

[53] S. Faez, P. Türschmann, H. R. Haakh, S. Götzinger, V. Sandoghdar, *Phys. Rev. Lett.* **2014**, *113*, 213601

[54] J. L. Goldfarb, *J. Heterocyclic Chem.* **2013**, *50(6)*, 1243-1263.

[55] M. Musavinezhad, A. Shkarin, D. Rattenbacher, J. Renger, T. Utikal, S. Götzinger, and V. Sandoghdar, *J. Phys. Chem. B* **2023,** *127(23)*, 5353-5359